\documentclass[aps,prl,twocolumn,superscriptaddress,nofootinbib,
nobalancelastpage,showpacs,nobibnotes]{revtex4}
\usepackage{epsfig}

\begin{document}
\title
{Axion, photon-pair mixing in models of axion dark matter }

\author{R. F. Sawyer}
\affiliation{Department of Physics, University of California at
Santa Barbara, Santa Barbara, California 93106}

\begin{abstract}

A system of light axions comprising a classical axion field, one candidate for dark matter, has an instability that would rapidly mix in photon pairs in a coherent fashion if the system were initially seeded by some tiny amount of such mixing. We develop equations that contain the mixing and at the same time incorporate enough quantum mechanics to eliminate the need for seeds.
Extending to many modes brings many interesting issues to the fore. For example, the argument of the inevitable logarithmic factor in the mixing time becomes reduced by many orders of magnitude; concerns concerning red-shifts are laid to rest, as are those related to lumpiness of the original axion state. We further see  that even the fully developed states of the electromagnetic field are completely non-classical in our solutions.
 \end{abstract}
 \maketitle

Cosmological models in which the dark matter is composed of light axions, in an essentially classical condensed state, have attracted attention recently \cite{peebles} - \cite{marsh}. Here we shall look again at the time evolution due to electromagnetic interactions of a piece of this matter, consisting of $N_a$ axions contained within a periodic box
of volume, $V$, and over a time interval somewhat less than the light travel time over the box. We assume a standard interaction,
$\mathcal{L}_I=g_\gamma a \vec E\cdot \vec B$, where $a$ is the axion field. Axion masses and couplings from literature are in the ranges, $g_\gamma=10^{-21} -10^{-22}  ~{\rm e V}^{-1}$  and
$10^{-22} {\rm eV}< m_a <10^{-4}$eV.

From early in the development of this subject, it has been known  \cite{wil}- \cite{fis} that in some regions there is an instability that could lead to exponential increase in mixing, with a term growing as $\exp[{r_{g} t}] $, where $ r_g\approx g_\gamma (\rho m_a)^{1/2}$ and $\rho =$[energy density]. 
In later literature \cite{l}-\cite{ll} possible consequences of this instability have been explored but its seeding has remained obscure. 

However beginning with a pure axion state there is no exponentially increasing photon number in the short term in our results. Instead there is a gestation time of order $r_g^{-1} \log ( \rho  m_a^{-4})$ during which little happens that is apparent, followed by a sudden near-complete, somewhat transitory, transformation of axions into photons. ``Sudden", here, means: on a time scale $r^{-1}$ much smaller than the logarithmic gestation time. 
This type of behavior can be designated a ``quantum break", a term that has gained currency in describing a genre of actual and conjectured phenomena in several areas: in condensed matter literature
describing, e.g. Bose condensates of atoms \cite{va}-\cite{va3} ; in polarization exchange processes in colliding photon beams  \cite{rfs1}-\cite{rfs4} ; in cosmology 
\cite{cos1}-\cite{cos3}. Finally, there is a close formal relation to ``fast neutrino flavor exchange" in the neutrino-sphere region in the supernova \cite{rfsx}-\cite{n12}, where the quantum term enabling the break is just the neutrino mass term. We mention the latter to emphasize that the underlying break dynamics are really not specific to Bose condensates, or even to bosons.
In each case the initial state is taken to be stable in a mean field theory (MFT). In each case there is a well defined break-time. For a case with a large number, $N$, of particles it is generally found that the time waiting for the break is of the order of $r_g^{-1} \log N$. But in the axion case we will find a greatly reduced value for the argument of the logarithm.

We define $c_{\bf q}^\dagger$, $c_{\bf q}$ to create and annihilate photons with momentum ${\bf q}$.
Next we write an effective interaction Hamiltonian that describes the mixing induced by the interaction in a lowest order calculation, and keep only terms that conserve momentum exactly, 
\begin{eqnarray}
H_{\rm eff}=& H_0+  V^{-1/2}\lambda \sum_{|{\bf q}|=m_a/2}  \Bigr [  b \,c_{\bf q}^\dagger  \,c_{\bf -q}^\dagger+
b^\dagger c_{\bf q}  \,c_{\bf -q} \Bigr ] 
\nonumber\\
 &+ \sum_{|{\bf q}|=m_a/2}\omega_p ( c_{\bf q}^\dagger  \,c_{\bf q}+c_{\bf -q}^\dagger  \,c_{\bf -q})\,,
\label{ham}
\end{eqnarray}

where 
\begin{eqnarray}
H_0={m_a \over 2} \sum [c_{\bf q}^\dagger  \,c_{\bf q}+c_{\bf -q}^\dagger  \,c_{\bf -q}]+m_a b^\dagger b \,,
\end{eqnarray}
and where $\lambda=g_\gamma  m_a^{1/2} $. Noting that $[H_0,H_{\rm eff}]=0$, we can set $H_0=0$ for simplicity.

We economized in notation by leaving out photon polarization indices in the above. A single breed, e.g., helicities =1 for both photons suffices. Including the state with helicities =-1 for both photons makes negligible effects on any results.

\subsection{ 1. Mean fields and quantum break}

For the first demonstration we select one particular photon pair direction $({\bf q,-q})$ in space and for now take $\omega_p=0$. 
For the most primitive definition of ``mean field approximation" we simply write the Heisenberg equations for our three operators, $ b, c_{\bf q} , c_{\bf - q}$, obtaining
$\dot c_{\bf q} =\lambda V^{-1/2} \,b\, c_{\bf -q}^\dagger\, $, etc.  Then we replace each of the operators therein by its expectation value; that is, we take the expectation of a product to be the product of expectations. Since we began with no photon field and $\langle c_{\bf \pm q}\rangle=0$, we see that the system stays exactly where it began. 

At the same time, if we explored the space with small initial $\langle c_{\bf \pm q}\ne 0\rangle$'s we would find exponentially increasing modes. Thus we categorize the original system as being in unstable classical equilibrium. Our mission is to calculate ``a quantum break time`` as discussed above.
We shall do this in two ways; first by creating an extended MFT based on operators that are quadratic in the original variables; second by just solving for the complete wave function, but in that case limited to a small number of axions. The basic agreement of the methods provides us with enough confidence in the extended MFT to proceed with predictions when, e.g., $N_a=10^{40}$.

We introduce three operators $X,Y,Z$,
\begin{eqnarray}
Z=b~;~Y= c_{\bf q}c_{\bf -q}~;~X=c^\dagger_{\bf q}c_{\bf q}+c^\dagger_{\bf -q}c_{\bf -q}\,,
\end{eqnarray}
with the effective Hamiltonian for the mode,
\begin{eqnarray}
H=\lambda V^{-1/2}[Z^\dagger Y+Z Y^\dagger] \,,
\label{ham2}
\end{eqnarray}
and introduce a scaled time variable $s=t \lambda V^{-1/2} N_a^{1/2}=t \lambda n_a^{1/2}$, where $n_a$ is the initial number density. 
In addition, we rescale the operators: $X=N_a x$, $Y=N_a y$, $Z=N_a^{1/2 }z$. Then the equations of motion for the operators $x,y,z$ under the interaction of the Hamiltonian (\ref{ham2}), are,
\begin{eqnarray}
i{d\over ds }z\, =y\,,
\nonumber\\
i{ d\over ds }y=(N_a^{-1}+x)\, z\,,
\nonumber\\
i{d\over ds } x= 2 (z y^\dagger - y z^\dagger) \,.
\label{mfes}
\end{eqnarray}

The MFT replaces each variable in these equations by its expectation in the medium. The initial conditions for our problem are $\langle z\rangle=1$; 
$\langle x \rangle=\langle y \rangle=0$.  The term $z$ in the  ${d\over ds }y$ equation has, in effect, one more power of $\hbar$ than the $x z$ term, having been produced from a final $c, c^\dagger$ commutator, and it enables the evolution starting from the pure axion state. The break that it induces is therefore identified as a 
``quantum" break. 
In the dashed curves of fig. 1 we show the time dependence of the residual axion fraction 
$\zeta(t)=N_a^{-1}\langle b^\dagger b\rangle$  derived from solutions of (\ref{mfes}) using the above initial condition, for a sequence of values of $N_a$, differing by a factor of two at each step. The equal spacings of the curves indicate a turnover time that increases as $\log N_a$. 

\begin{figure}[h] 
 \centering
\includegraphics[width=2.5 in]{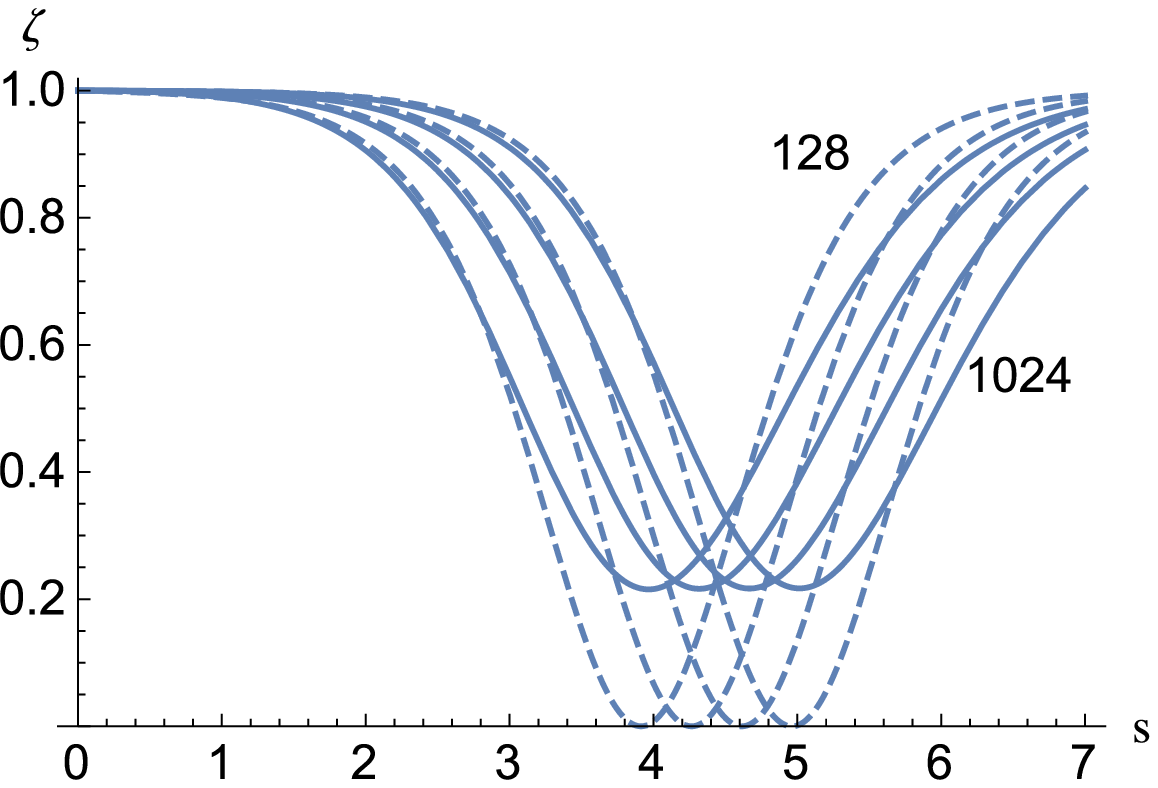}
 \caption{ \small }
Evolution in scaled time $s$. $\zeta$ is the persistence probability for an axion. The solid curves are the result of the solution of the Schrodinger equation for $N_a=128, 256, 512,1024$ going from left to right. The dashed curves are the mean-field solutions for the same values of $N_a$.
\end{figure}
For the relatively small values of $N_a$ used in these plots we can instead do a complete quantum calculation of the wave function, based on the Hamiltonian (\ref{ham}) (but still with a single value of ${\bf q}$), beginning with the pure axion state. This requires the solution of $3(N_a+1)$ simultaneous, coupled, first-order, linear equations. At laptop Mathematica level we can solve the system for values of $N_a \le 1000$. Results are shown as the solid curves of fig. 1, for the same set of $N_a$ as used in the MFT model. The agreement of the two calculations is good only up to the inflection point midway through the
break.  On the other hand, the equal spacings of the minima, as we repeatedly double $N_a$, are remarkably similar in the two calculations. The bounce at about $\zeta=.2$ for the complete quantum case, while the MFT result goes all the way to zero, is mysterious.  There is a further qualitative difference in that the quantum solution does not return to $\zeta=1$ in the finite $N_a$ solutions. Indeed, when extended to longer times, it appears to experience very irregular jagged oscillations around 
a value $\zeta=.6$, while the mean field solutions are periodic. In any case, all of our arguments for physical relevance will be based on the location of the first break and a large mixing at time,

\begin{eqnarray}
T\approx (\lambda n_a^{1/2})^{-1}  \log_{10} [N_a]\,,
\label{time}
\end {eqnarray}
 where the base 10 is a rough fit to the spacings shown in fig. 1.
In fig. 2 we show the continuation of the $\log N_a$ behavior in the MF solution for larger $N_a$,  successively higher by factors of 100.\begin{figure}[h] 
 \centering
\includegraphics[width=2.5 in]{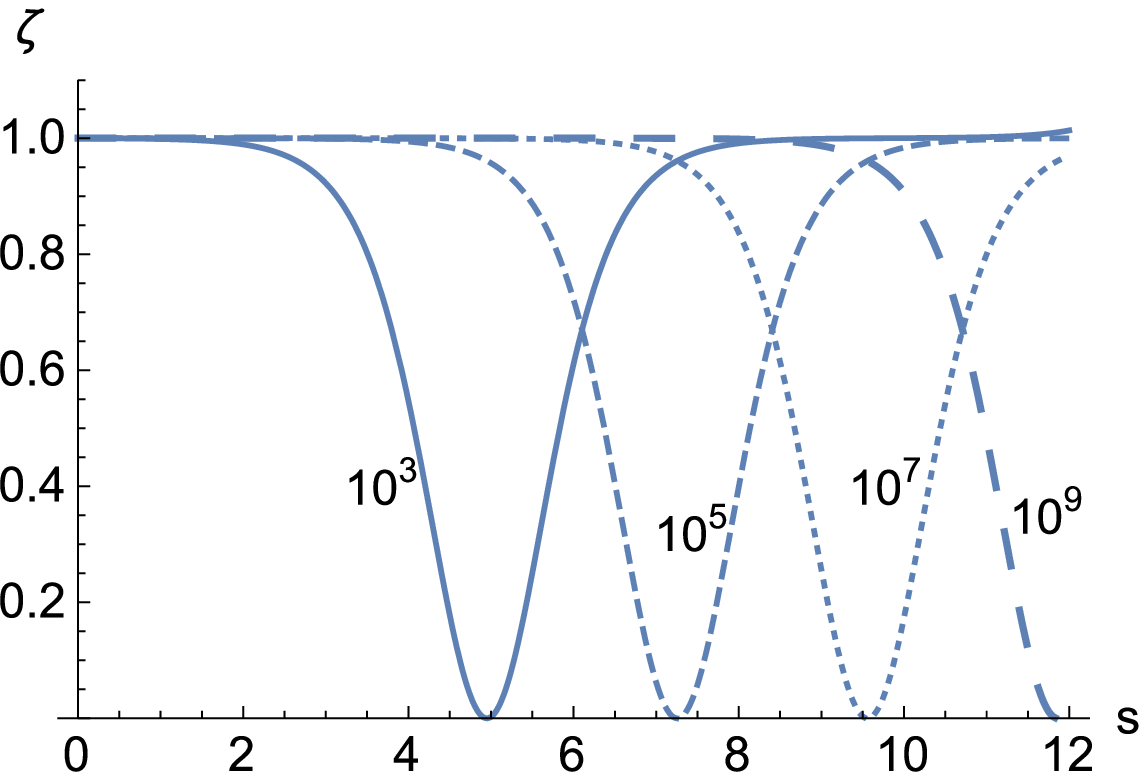}
 \caption{ \small }  Mean-field behavior, as in the dotted cubes in fig1, but for values of $N_a=10^3, 10^5,10^7, 10^9$
 \label{ fig. 2}
\end{figure}
Note that the ``break" of each these plots has exactly the same apparent shape in scaled time. The duration $ \Delta T$ of the break itself  is
 $\Delta T\sim T/ \log_{10}[N_a]$, the logarithmic part of the total time going entirely into the nearly imperceptible simmering stage.

\subsection{2.Many angles}

An issue that was implicitly raised in arriving at (\ref{time}), was the fact that we chose one direction in space for the final photon pair. The wave-function for the system, in its quantum simmering phase before the break, is perfectly able to run away in many directions simultaneously. The calculation in a box of side $L$ puts a limit on the number of allowable directions for the photons, which at the order-of magnitude level is $N_d\approx m_a^2 L^2 $. 
In the MF approach we introduce the notations $c_k ,\, \tilde c_k$ as the respective annihilation operators for photons with momenta ${\bf q_k},{\bf - q_k}$
and define operators,
\begin{eqnarray}
Z=b~;~Y_k=  c_{k }\tilde c_{k}~;~X_{k}=c^\dagger_{k} c_{k}+\tilde c^\dagger_{k} \tilde c_{k}\, ,
\end{eqnarray}
with the Hamiltonian
\begin{eqnarray}
H=g (Z \sum_k^{N_d}  \lambda_k Y_k^\dagger+Z ^\dagger \sum_k^{N_d}  \lambda_k Y_k)\,.
\label{multi}
\end{eqnarray}

The equations for the rescaled $x_k,y_k,z$ are
\begin{eqnarray}
i{d\over ds }z\, =\sum_j^{N_d} y_k\,,
\nonumber\\
i{ d\over ds }y_k=(N^{-1}+x_k)\, z + \Bigr \{  \Bigr [{2 \bar \omega \, k \over N_d  } \Bigr ]\,y_k\,\Bigr\}
\nonumber\\
i{d\over ds } x_k= 2 (z y_k^\dagger - y_k z^\dagger) \,.
\label{mf}
\end{eqnarray}
where the term in the curly bracket will be explained and used in a later section. 
We now can verify that random changes at the 20\% level of individual couplings, away from from a universal value $\lambda_k=1$ , make almost no difference to the axion disappearance plot.

Ideally we would have checked the agreement
of the mean-field approach with the complete solutions over a wide range of $N_a$, and $N_d$, but in the complete case we could only afford $N_d$=2. In fig. 3 we show the comparison of the results of MFT calculation of (\ref{mf}) to those of the complete quantum calculation of the Schrodinger wave-function.

\begin{figure}[h] 
 \centering
\includegraphics[width=2.5 in]{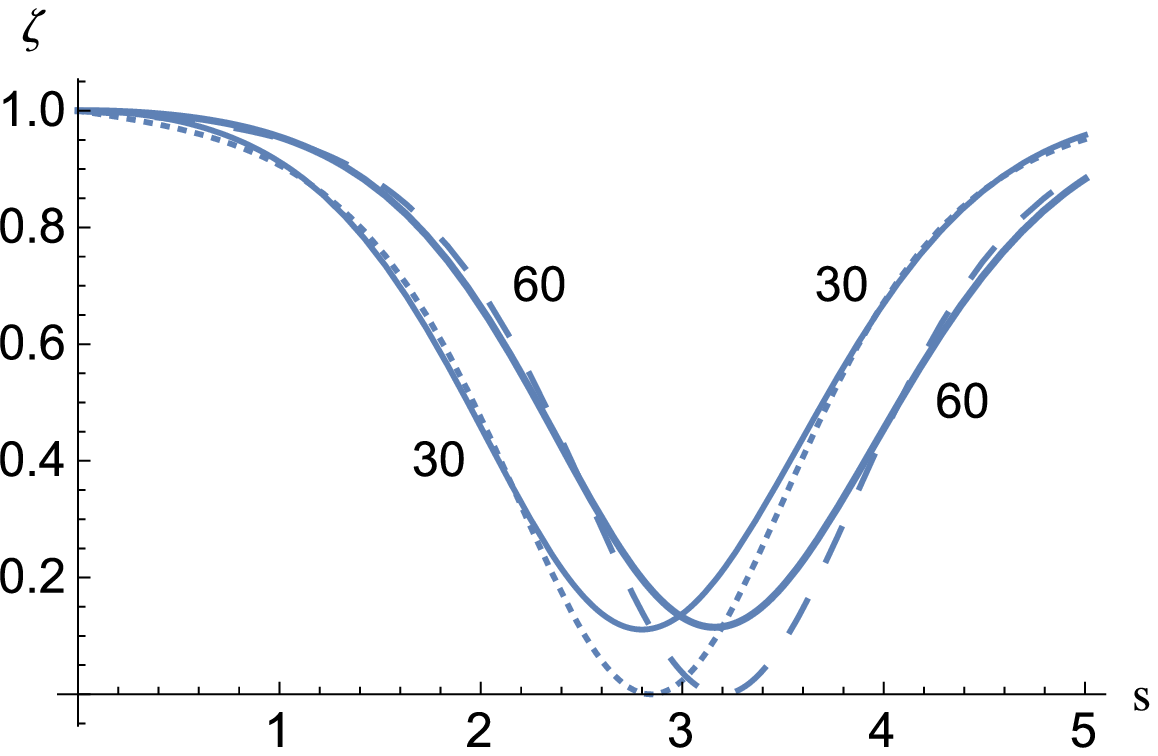}
 \caption{ \small }
Solid curves are the complete quantum solution for persistence probability $\zeta$ in the case of two groups (angles) for $N_a=30, 60$. Dashed curves are
the two group mean-field result. 
 \label{ fig. 1}
\end{figure}
The agreement of the mean-field approach with the complete solution is even improved somewhat over the single angle case. This emboldens us to use the MF approach when $N_a$ and $N_d$ are both large.
Our conclusion after many MF solutions for different values of $N_a$ and $N_d$ is that the turnover time, in our basic unit $T_0=\lambda^{-1} n_a^{-1/2} $,  is now approximately
$T\sim T_0 \log [N_a/ N_d]$, so long as $N_d<<N_a$. The effect of the additional final channels is the mitigation of the logarithmic factor. Putting in the above estimate of $N_d$, we now have $T\sim T_0 \log [ L m_a^{-2} n_a]$. 

We go back to the beginning, for a moment, and relax the decision to restrict our set of states to those that exactly conserve energy.  We add in the effects of
$N_t$ new  2$\gamma$ modes, each with equal additions $q_T$ to the two transverse photon momenta, constrained by $q_T^2/m_a<m_a $ in order to maintain near-coherence. Counting these states we find $N_t=L q_T^{\rm max}=L  m_a$. Then the final result for the turnover time is of order,
\begin{eqnarray}
T\sim T_0 \log \bigr [{N_a \over N_d  N_t}\bigr ]  \approx T_0 \log [n_a m_a^{-3} ]\,.
\label{log}
 \end{eqnarray}.

\subsection{3. Inhomogeneity and red shift}
All of the above was based on a coupling, turned on at $t=0$, of photons to a pure axion state that occupies a volume $V$ in space. Here the early dark era is an appealing choice of venues, because any development of coherent correlations of axions with the electromagnetic field before recombination would have been cut off by Compton interactions.  Distributions were still fairly homogeneous at the beginning of the era. So we envision a turn-on time to be set at a recombination time that is simultaneous for everybody, though of course the different regions weren't in touch at that time. But there were already density fluctuations at some level over the path lengths involved in our mixings, and we need to investigate their possible effects.

We have written a program to do this (in 1+1 D), discretized by dealing with a chain of $N_b$ boxes in a row, each one passing on its information repeatedly after small increments in time, to the box immediately to the right, for the right-moving amplitudes and the one to the left for the left-moving amplitudes.  We use the equations (\ref{mf}) to formulate this problem where the index, k, which had stood for the photon direction in 3D, now identifies a box. 
The calculation using these effective packets gives the very nearly the same answer for the transition time as does the previous single box calculation, with the differences ascribable to the discretization in space. As a bonus we find that if we introduce, for each site, random variations in the local axion density at the 
10-20\% level (with consequent changes in local effective coupling) the long term behavior is altered hardly at all. This result is related to well-understood line-narrowing mechanisms referred to as ``motional narrowing" in NMR \cite{mot}, or a precursor, N. F. Ramsey's "split-field" technique for reducing line width in a molecular beam experiment \cite{ram}.

The red-shift issue, which has been mentioned by a number of authors, is concerned with cases in which the maximally red-shifted photons received at our location at the turnover time $T$ is greater than a ``width", as estimated from $T^{-1}$. Therefore, the argument goes, the developing mixing will be snuffed out before it becomes significant. One over-simplification of the above is clear, since if we divide into $N_s$ time boxes, where $N_s$ is large,  then in the short first time interval after turn-on, in the first box, our quantum calculation can turn its crank and produce just as much photon amplitude in a much wider cluster of states 
$\pm q$, in the entire range $|q|> m_a[1+T \dot a(t) /a(t)] $, where $a(t)$ is the scale factor. We can take the Hubble rate, $\dot a(t) /a(t)$, as constant over our time span   So from this range we select the momentum region that is blue-shifted (in its local system) from the usual resonance band by just enough to arrive at the end (our place) on center.  And the photons in this little band don't get very out-of phase with each other in moving into the next-door box nearer to us. They move in happy to stimulate more emission into their various momentum states. In addition, they have now moved a little closer to the ``resonance", in the frame of the second box, since in the now local system they do not appear as blue-shifted as they were in the previous step; and so forth, working their way towards us. We have written another simulation, this a time-box simulation, that implements the above. The results are preliminary but appear to indicate no important red shift effect.

\subsection{Relation to other work}
 In our view it is dubious to think of this problem as single-mode and essentially classical, as it is usually cast in the rapidly growing literature. For the moment dropping the red-shift questions, which would have forced us to multi-mode considerations in any case, we see the following issues:
 
 1) The Mathieu equation approach to the linear perturbation region (ref. \cite{hz} and the references contained therein), an essentially classical approach, requires a non-vanishing initial value for the time derivatives of the $E \&M$ fields (often thought of as a ``vacuum fluctuation"). Then its early time behavior can be characterized by an increasing exponential based on a Lyapunov exponent.  But our set (\ref{mfes}), as is, very gradually and automatically seeds the system. The fields begin at zero with zero derivative and increase as $t^2$ in the small $t$ region. There is an intermediate time interval over which a Lyapunov exponent does a major part of the work. And by best-fitting the putative ``vacuum fluctuation" seed parameters in the classical model we can get a not too bad fit to our solutions for the axion retention probability, except in the early times, but not as good as the fits shown in fig. 1 for the early era based on no arbitrary parameters.
  
 2) We must emphasize that in the quantum case the individual expectations of the E\&M operators, $c_q, c_{-q}$ are zero through-out the evolution, whereas in the mostly classical view, they carry the evolving classical field. In our multi-angle simulation beginning with homogeneous and isotropic system of axions, how then
 could we have obtained this homogeneous and isotropic photon state, since there is no such classical solution? The answer is that we obtained instead a quantum superposition of states of different directions, with the quantum state having no preferred direction. The physical content is isotropic, but it is not even approximately represented by a classical field. Or maybe it is better to say ``the expectation of the bilinear of this field is not the bilinear of the expectations."  If we had assigned a classical field to each ray, and then superposed those, then as we approached a continuum, the summed field would have been zero, and almost zero had we stopped short of a continuum limit. 
 
 But if sitting here at some local point in space with our photon counter we waited to see a bunch come by we would see the same pulse at our time T (or better, at the same red-shift as us), as would another observer over the universe at the same red-shift. We are not here aspiring to do quantum cosmology here; we are just pointing out the difference between our results and classical assumptions about how things work in this idealized system that started as pure axion field.  
 \subsection{5. Discussion}

We have given an argument that axions in dense clouds, with the usual form of electromagnetic coupling, can mix strongly in a coherent way with photon pairs.  Do current models of axion dark matter ever yield the combination of axion density and cloud size that makes our calculation relevant? 
First we look back in time at the evolution of a cloud when the DM density was nearly uniform, but where this density was higher by a factor of $z^3$ where $z $ is the red shift. For $z=10^3$, the beginning of the ``dark era" (just post recombination) the energy density of dark matter is $\rho \approx$ .04 (eV)$^4$.  For the case  $g=10^{-21} $ eV$^{-1}$
the basic distance scale for turn-over is a few light-years times the logarithmic factor. This era has low electron density,
and a pure axion initial state is plausible at some point in time; prior coherent mixings with photons were suppressed by the high free electron density
before recombination. The plasma frequency in this dark era could be as low as $\omega_p \approx 10^{-11} $ eV (taking H ionization of 10$^{-5}$). The axion cloud would then be a venue for the application of the results of this paper only in a domain $m_a>10^{-11}$ eV. For this case we estimate the logarithmic factor from the the number of photon states that enter, as roughly enumerated in (\ref{log}), and this brings us to an estimate of 100 LY , far less than the horizon size at that time.

So perhaps there could be such an event. But the author would be the first to acknowledge that, if for no other reason than that of the absurdly small numbers of states used in the supporting simulations, everything here must be reworked, either with more powerful computational resources, or with better analytical approaches. Of course, if the axion-coupling to photon were to produce such mixings, then when they get large, as time goes on, the photon component will interact enough with other stuff in the surroundings to break the spell, leading to interesting observable phenomena. Or if the final state of our mixing is a 50-50 mixture, then gravity would probably start messing it up, since it acts so differently on the different components.
Axion stars \cite{marsh2}-\cite{star4} with enormously higher densities could be a more promising site for coherent mixing. However, in more complicated geometries it may be difficult to follow the coherent development of axion photon-systems. 

The present paper contains the following new material:

1. A modified mean-field approach that gives the required quantum break, without arbitrary assumptions about vacuum fluctuations turning into classical seed fields. Such a break is necessary for the development of large mixing.

2. A complete solution of the Schrodinger equation for values $N_a \le1000 $ that supports the modified mean-field approach.

3.  A demonstration that including many modes of the electromagnetic field greatly reduces the argument of the inevitable logarithm.

4. An explanation as to how, beginning with an interaction in which many modes of the photon field couple to the axion field, we avoid the perceived ``red shift" problem. with coherence over very long times.

5. An account of calculations that indicate that when a clumpy axion distribution is encountered by photons {\it en route} to their destinations is not disruptive to the turn-over phenomenon.

The author thanks Mark Srednicki for a critical observation and Alessandro Mirizzi for a very useful comments.
  
\end{document}